\documentclass[aps,onecolumn]{revtex4}

\usepackage[dvips]{graphicx}
\usepackage{color}
\begin{document}

\title{ {\tt {\small \begin{flushright}
{\tt CERN-PH-TH/2013-197\\ 
MAN/HEP/2013/18\\ 
August 2013 } 
\end{flushright} } }
Lepton Dipole Moments in Supersymmetric Low-Scale Seesaw Models}

\author{Amon Ilakovac$^{\,a}$, Apostolos Pilaftsis$^{\,b,c}$ and Luka
  Popov$^{\,a}$\vspace{2mm}}

\affiliation{
${}^a$University of Zagreb, Department of Physics,
  Bijeni\v cka cesta 32, P.O. Box 331, Zagreb,
  Croatia\vspace{1mm}\\
${}^b$CERN, Department of Physics, Theory Division, CH-1211 Geneva 23,
  Switzerland\vspace{1mm}\\
${}^c$Consortium for Fundamental Physics,
  School of Physics and Astronomy, University of Manchester,
  Manchester M13 9PL, United Kingdom
}

\begin{abstract}
\noindent
We study the anomalous magnetic and electric dipole moments of charged
leptons  in supersymmetric low-scale  seesaw models  with right-handed
neutrino superfields.   We consider a minimally  extended framework of
minimal supergravity,  by assuming  that CP violation  originates from
complex soft SUSY-breaking bilinear and trilinear couplings associated
with   the  right-handed   sneutrino  sector.    We~present  numerical
estimates  of the  muon  anomalous magnetic  moment  and the  electron
electric dipole  moment (EDM), as  functions of key  model parameters,
such as the Majorana mass scale~$m_N$ and~$\tan\beta$.  In particular,
we  find that  the contributions  of the  singlet heavy  neutrinos and
sneutrinos  to the  electron EDM  are  {\em naturally}  small in  this
model, of order $10^{-27}$--$10^{-28}~e\; {\rm cm}$, and can be probed
in the present and future experiments.

\medskip

\end{abstract}

\maketitle

\medskip

\setcounter{equation}{0}

\section{Introduction}

The  anomalous magnetic  dipole  moment (MDM)  of  the muon,  $a_\mu$,
constitutes a high precision observable extremely sensitive to physics
beyond  the  Standard  Model  (SM).  Its  current  experimental  value
$a_\mu^{\rm exp}=(116592089  \pm 63)\times 10^{-11}$  differs from the
SM  theoretical  prediction  $a_\mu^{\rm  SM}=(116591802\pm  49)\times
10^{-11}$ by~\cite{PDG2013}
\begin{equation}
\Delta a_\mu\ \equiv\ a_\mu^{\rm exp} - a_\mu^{\rm SM} \ =\ (287\pm
80)\times 10^{-11}\, .
\end{equation}
Evidently,  the  deviation  $\Delta  a_\mu$ is  at  the  $3.6\,\sigma$
confidence  level (CL)  and has  therefore been  called the  {\it muon
  anomaly}.  Consequently,  an important constraint  on model-building
is derived by requiring  that new-physics contributions to~$a_\mu$ are
smaller than $\Delta a_\mu$.

Likewise, the electric dipole moment  (EDM) of the electron, $d_e$, is
a  very sensitive  probe for  CP violation  induced by  new  CP phases
beyond  the  SM.   The present  upper  limit  on  $d_e$ is  quoted  to
be~\cite{PDG2013,JH2011,MJ2013}
\begin{equation}
  \label{deUB}
d_e\ <\ 10.5\times 10^{-28}\ e\; {\rm cm}\; .
\end{equation} 
Future projected  experiments utilizing paramagnetic  systems, such as
Cesium, Rubidium  and Francium, may extend the  current sensitivity to
the      $10^{-29}$--$10^{-31}\     e\;      {\rm     cm}$~level~(see,
e.g.~\cite{MJ2013}   and  references   therein).   In   the   SM,  the
predictions  for  $d_e$  range   from  $10^{-38}\  e\;  {\rm  cm}$  to
$10^{-33}\ e\; {\rm  cm}$, depending on whether the  Dirac CP phase in
the light neutrino mixing is zero or not~\cite{MPAR2005}.  Clearly, an
observation  of   a  non-zero   value  for~$d_e$,  much   larger  than
$10^{-33}\ e\;  {\rm cm}$,  would signify CP-violating  physics beyond
the SM.

As  an  archetypal  model   of  New  Physics,  the  so-called  Minimal
Supersymmetric Standard Model (MSSM) is of great interest.  In general,
models of  softly broken supersymmetry (SUSY) at  the 1--10~TeV scale,
such as the MSSM, can account for the gauge hierarchy problem, predict
rather  accurate  unification  of   gauge  couplings  near  the  Grand
Unification  Theory  (GUT)  scale,  naturally explain  the  origin  of
spontaneous symmetry breaking of the SM gauge group and predict viable
candidates for solving  the Dark Matter (DM) problem  in the Universe.
For a recent review, see~\cite{DIK2010}.

To account for the observed light neutrino masses and mixings, we will
consider  SUSY  extensions~\cite{DV} to  models  with low-scale  heavy
neutrinos  \cite{LSHN1,LSHN2,LSHN3,LSHN4}.    Specifically,  the  MSSM
extended  with low-scale right-handed  neutrino superfields,  which we
denote   hereafter   as   $\nu_R{\rm   MSSM}$,   predicts   additional
contributions to  charged lepton flavour violation (CLFV)  that do not
exist in models with high-scale heavy neutrinos and are independent of
the soft SUSY-breaking mechanism~\cite{IP_SLFV}.  It is interesting to
note   that    in   the   $\nu_R{\rm    MSSM}$,   $Z$-boson   penguins
\cite{IP_SLFV,Zamp_MSISM} and box diagrams \cite{IPP2013} dominate the
amplitudes of processes, such  as lepton$\;\to 3\;$leptons and $\mu\to
e$  conversion, whereas photon-penguin  LFV diagrams  are sub-dominant
and become only relevant to models with ultra-heavy neutrinos close to
the GUT  scale~\cite{GUTLFV}.  In particular, our  recent analysis has
shown~\cite{IPP2013}  that  a  significant  region of  the  $\nu_R{\rm
  MSSM}$ parameter space exists for which the branching ratios of CLFV
processes  are  predicted to  be  close  to  the current  experimental
sensitivities, despite the fact  that the soft SUSY-breaking scale has
been pushed  to values  higher than $1$~TeV,  as a consequence  of the
discovery of a  SM-like Higgs boson at the  CERN Large Hadron Collider
(LHC)~\cite{Hmass}  and  the existing  non-observation  limits on  the
gluino   and  squark   masses  that   were  also   deduced   from  LHC
data~\cite{m-gtqt}.

It is  therefore of particular  interest to investigate  here, whether
the effects of low-scale heavy  neutrinos and their SUSY partners, the
sneutrinos, contribute  in a relevant  manner to other  high precision
observables, such as  the muon anomalous MDM $a_\mu$  and the electron
EDM $d_e$.  We believe that the announced higher-precision measurement
of~$a_\mu$ by a  factor of $4$ in the  future Fermilab experiment E989
\cite{LeeR,GV2012}  and  the  expected  future  sensitivities  of  the
electron  EDM down  to  the level  of  $\sim 10^{-31}\  e\; {\rm  cm}$
\cite{MJ2013}, renders such an investigation both very interesting and
timely.

Most studies on lepton dipole moments have been devoted to SUSY models
realizing       a        high       scale       seesaw       mechanism
\cite{EHRS02_SUSYdip,IM03_SUSYdip,Peskin,Dedes2007}.  Here  instead, we consider
the   $\nu_R{\rm  MSSM}$   which   provides  potentially   significant
contributions to lepton dipole  moments due to low-scale neutrinos and
sneutrinos, as well as new  sources of CP violation. In particular, an
interesting possibility  emerges if  there exists CP  violation beyond
the  SM which is  sourced from  the singlet  sector of  the $\nu_R{\rm
  MSSM}$.  This  new CP  violation may originate  from a  complex soft
trilinear sneutrino parameter $A_\nu$, or from a complex soft bilinear
parameter  $B_\nu$.   In addition,  one  may  have  new CP-odd  phases
residing  in the  $3\times  3$ neutrino  Yukawa-coupling matrix  ${\bf
  h}_\nu$. Assuming that these are the only additional non-zero CP-odd
phases  in the $\nu_R{\rm  MSSM}$, we  find that  the electron  EDM is
testable, but  naturally small, typically of  order $10^{-27}~e\; {\rm
  cm}$, thereby avoiding to some  extent the well-known problem of too
large CP violation, from which SUSY  extensions of the SM, such as the
MSSM (see, e.g.~\cite{ELP}), usually suffer.

The outline  of the paper is  as follows. In Section  II, we introduce
our conventions and notation for the lepton dipole moments, as well as
describe the  new sources of CP  violation that we  are considering in
the $\nu_R{\rm  MSSM}$.  Section~III presents  our numerical estimates
for  the lepton dipole  moments $a_\mu$  and $d_e$.   To this  end, we
specify our  input parameters, including the  neutrino Yukawa matrices
adopted  in  our  numerical   analysis.   Section  IV  summarizes  our
conclusions.  Technical details  pertinent to the lepton-dipole moment
formfactors are given in Appendix \ref{f3.4and3.5}.

\section{Magnetic and electric dipole moments}
\setcounter{equation}{0}

The anomalous MDM and EDM of a charged lepton $l$ can be read off from
the Lagrangian~\cite{Branco}:
\begin{equation}
  \label{l_effL}
{\cal L}\ =\ \bar{l}\,\Big[\gamma_\mu(i\partial^\mu + e A^\mu) 
  - m_l - \frac{e}{2m_l} \sigma^{\mu\nu} (F_l+iG_l\gamma_5) 
                                       \partial_\nu A_\mu\Big]\, l\; .
\end{equation}
In the  on-shell limit  of the photon  field $A^\mu$, the  form factor
$F_l$ defines  the anomalous MDM  of the lepton $l$,  i.e.~$a_l \equiv
F_l$, whilst the  form factor $G_l$ defines its  EDM, i.e.~$d_l \equiv
eG_l/m_l$.  Given  that the  general form-factor decomposition  of the
photonic transition amplitude is given by~\cite{IPP2013}
\begin{equation}
i{\cal T}^{\gamma ll}\ =\
 i\frac{e \alpha_w}{8 \pi M_W^2}
 \Big[ (G_\gamma^L)_{ll} i\sigma_{\mu\nu}q^\nu P_L
 + (G_\gamma^R)_{ll} i\sigma_{\mu\nu}q^\nu P_R \Big]\; ,
\end{equation}
the anomalous  MDM $a_l$ and  the EDM $d_l$  of a lepton $l$  are then
respectively determined by
\begin{eqnarray}
\label{aldl_nRS}
 a_l &=& \frac{\alpha_w m_l}{8\pi M_W^2} \Big[ (G_\gamma^L)_{ll} + 
        (G_\gamma^R)_{ll} \Big]\ ,\\
 d_l &=& \frac{e \alpha_w}{8\pi M_W^2} i \Big[ (G_\gamma^L)_{ll} - 
         (G_\gamma^R)_{ll} \Big]\ .
\end{eqnarray}
Here and in the following, we adopt the notation for the couplings and
the form-factors established in~\cite{IPP2013}.

At the  one-loop level, the  EDM $d_l$ of  the lepton vanishes  in the
MSSM with universal soft SUSY-breaking boundary conditions and no soft
CP  phases,   adopting  the   convention  of  a   real  superpotential
Higgs-mixing  parameter~$\mu$~\cite{Peskin}.  This  result  also holds
true, even in extensions of the  MSSM with heavy neutrinos, as long as
the sneutrino sector is universal and CP-conserving as well.

As a minimal departure of the above universal scenario, we assume here
that {\em only}  the sneutrino sector is CP-violating,  due to soft CP
phases in the bilinear and trilinear soft-SUSY breaking parameters:
\begin{eqnarray}
  \label{CPBnu}
{\bf b}_\nu &\equiv& {\bf B}_\nu {\bf m}_M \ =\ B_0 e^{i\theta} m_N {\bf 1}_3\;,\\ 
  \label{CPAnu}
{\bf A}_\nu &=&  {\bf h}_\nu\, A_0 e^{i\phi}\; ,
\end{eqnarray}
where $B_0$ and $A_0$ are real parameters determined at the GUT scale,
$m_N$ is a real parameter inputed at the scale $m_N$, and $\theta$ and
$\phi$ are  physical, flavour blind CP-odd phases.  In addition, ${\bf
  h}_\nu$ is the $3\times 3$ neutrino Yukawa matrix to be specified in
the next section.   The soft SUSY breaking terms  corresponding to the
${\bf b}_\nu$ and ${\bf A}_\nu$ are obtained from the Lagrangian terms
\begin{equation}
 - ({\bf A}_\nu)^{ij} \tilde{\nu}^c_{iR} (  h^+_{uL}
  \tilde{e}_{jL} - h^0_{uL} \tilde{\nu}_{jL})\\  
\end{equation}
and 
\begin{equation}
({\bf b}_\nu {\bf m}_M)_{ii} \tilde{\nu}_{Ri}\tilde{\nu}_{Ri}\; ,
\end{equation}
respectively. Correspondingly, $\tilde{\nu}^c_{iR}$, $\tilde{e}_{jL}$,
$h^+_{uL}$  and  $h^0_{uL}$  denote  the heavy  sneutrino,  selectron,
charged Higgs  and neutral Higgs fields.  The  $O(3)$ flavour symmetry
of the model  for the heavy neutrinos assures  that the heavy neutrino
mass matrix ${\bf m}_N$ is proportional to the unit matrix ${\bf 1}_3$
with eigenvalues $m_N$, up to small renormalization-group effects.  To
keep things simple, we also  assume that the $3\times 3$ soft bilinear
mass  matrix ${\bf  b}_\nu$ is  proportional to  ${\bf 1}_3$.   In the
standard  SUSY seesaw  scenarios  with ultra-heavy  neutrinos of  mass
$m_N$, the CP-violating sneutrino  contributions to electron EDM $d_e$
scale  as   $B_0/m_N$  and  $A_0/m_N$  at  the   one-loop  level,  and
practically decouple for heavy-neutrino  masses $m_N$ close to the GUT
scale.  Hence,  sizeable effects on  $d_e$ should only be  expected in
low-scale  seesaw  scenarios, in  which  $m_N$  can become  comparable
to~$B_0$ and~$A_0$.

Following  the  conventions  of~\cite{IPP2013},  the  $12  \times  12$
sneutrino mass matrix may be cast into the $4\times 4$ block form:
\begin{eqnarray}
   \label{M2snu1}
{\bf  M}^2_{\tilde{\nu}}
 &=&
 \left(
 \begin{array}{cccc}
 {\bf  H}_1 & {\bf  N} & {\bf  0} & {\bf  M} \\
 {\bf  N}^\dagger & {\bf  H}_2^T & {\bf  M}^T & {\bf b}_\nu^\dagger \\
 {\bf  0} & {\bf  M}^* & {\bf  H}_1^T & {\bf  N}^* \\
 {\bf  M}^\dagger & {\bf b}_\nu & {\bf  N}^T & {\bf  H}_2
 \end{array}
 \right)\; .
\end{eqnarray}
The entries of ${\bf M}^2_{\tilde{\nu}}$ are expressed in terms of the
$3\times 3$ matrices:
\begin{eqnarray}
  \label{M2snu2}
{\bf  H}_1 &=& {\bf  m}^2_{\tilde{L}} + {\bf  m}_D {\bf  m}_D^\dagger
+ \frac{1}{2} M_Z^2 \cos 2\beta\; ,\nonumber\\
{\bf  H}_2 &=& {\bf  m}^2_{\tilde{\nu}} + {\bf  m}_D^\dagger {\bf
  m}_D + {\bf  m}_M {\bf  m}_M^\dagger\; ,\\[2mm]
{\bf  M} &=& -\frac{v_2}{\sqrt{2}}{\bf  A}^\dagger_\nu - \mu {\bf  m}_D \cot\beta\; ,\nonumber\\[2mm]
{\bf  N} &=& {\bf  m}_D {\bf  m}_M\; .\nonumber
\end{eqnarray}
Here  ${\bf  m}^2_{\tilde{L}}$,  ${\bf m}^2_{\tilde{\nu}}$  and  ${\bf
  A}_\nu$ are $3\times 3$  soft SUSY-breaking matrices associated with
the  left-handed  slepton doublets,  the  right-handed sneutrinos  and
their trilinear  couplings, respectively.   We note that  the bilinear
soft    $3\times   3$~matrix~${\bf    b}_\nu$    was   neglected    in
Ref.~\cite{IPP2013},  where the  authors tacitly  assumed that  it was
small   compared   to   the   other  soft   SUSY-breaking   parameters
in~(\ref{M2snu1}).  Here, we take this term into account, but restrict
the size of the universal bilinear mass parameter $B_0$, such that the
sneutrino masses remain always positive and hence physical.

The generation of a non-zero EDM $d_e$ results from the soft sneutrino
CP-odd phases  $\theta$ and $\phi$,  as well as from  complex neutrino
Yukawa couplings ${\bf h}_\nu$. All these CP-odd phases are present in
the  photon  dipole   form  factors  $G_{ll\gamma}^{L,\tilde{N}}$  and
$G_{ll\gamma}^{R,\tilde{N}}$,  whose  analytical  forms may  be  found
in~\cite{IPP2013}.   In  fact, we noticed that $d_e$ may be generated  by
products of vertices that are not relatively complex conjugate to each
other, such as~\cite{CPNt}
\begin{equation}
  \label{CPNt}
\Delta^{LR}_{\rm CP}\ =\ \tilde{B}^{L,1}_{l k A} \tilde{B}^{R,1*}_{l k A}\: +\:
 \tilde{B}^{L,2}_{l k A} \tilde{B}^{R,2*}_{l k A}\; ,\qquad
\Delta^{RL}_{\rm CP}\ =\ \tilde{B}^{R,1}_{l k A} B^{L,1*}_{l k A}\: +\:
 \tilde{B}^{R,2}_{l k A} \tilde{B}^{L,2*}_{l k A}\; .
\end{equation} 

In the exact supersymmetric  limit of softly-broken SUSY theories, the
anomalous MDM (as well as EDM) operator is forbidden, as a consequence
of the  Ferrara and Remiddi no-go  theorem~\cite{FerRem}.  The theorem
can  be   verified  for   every  particle  and   its  SUSY-counterpart
contribution   to  the   anomalous  MDM   $a_\mu$.   Besides   the  SM
contribution,  there   are  three  additional   contributions  in  the
$\nu_R$MSSM,    which    originate    from:    (i)~heavy    neutrinos,
(ii)~sneutrinos  and~(iii)  soft  SUSY-breaking  parameters.   In  the
supersymmetric limit, the  latter contribution~(iii) vanishes.  In the
same limit, the heavy neutrino and sneutrino contributions read:
\begin{eqnarray}
(G_{\gamma}^{ll})^N &\to& \frac{7}{6} B_{lN_a} B_{lN_a}^*\; ,
\nonumber\\
(G_{\gamma}^{ll})^{\tilde{N}} &\to& -\ \frac{7}{6} B_{lN_a} B_{lN_a}^*\; , 
\end{eqnarray}
where $B_{lN_a}$  are the lepton-to-heavy neutrino  mixings defined in
the  first article  of Ref.~\cite{LSHN4}  and in  Ref.  \cite{IP_NPB}.
Obviously,        the         sum        $(G_{\gamma}^{ll})^N        +
(G_{\gamma}^{ll})^{\tilde{N}}$   vanishes,   thereby  confirming   the
Ferrara--Remiddi theorem.

In  the   MSSM,  the  leading  contribution  to   $a_l$  behaves  as
\cite{TM1995,Sto06}.
\begin{eqnarray}
  \label{al_approx}
a_l^{\rm MSSM} &\propto& \frac{m_l^2}{M_{\rm SUSY}^2}\, \tan\beta\;
\mbox{sign}(\mu M_{1,2})\; , 
\end{eqnarray}
where  $M_{\rm SUSY}$  is  a typical  soft  SUSY-breaking mass  scale,
$\tan\beta=v_2/v_1$  is   the  ratio  of  the   neutral  Higgs  vacuum
expectation  values,  and  $M_{1,2}$   are  the  soft  gaugino  masses
associated  with the  ${\rm  U}(1)_{\rm Y}$  and  ${\rm SU(2)}$  gauge
groups, respectively.   As we will see  in the next  section, the MSSM
contribution~(\ref{al_approx})  to  $a_\mu$  remains dominant  in  the
$\nu_R$MSSM as well.

From~(\ref{al_approx}) and~(\ref{aldl_nRS}), one naively expects $d_l$
to behave at the one-loop level as
\begin{eqnarray}
  \label{di_approx}
d_l^{\rm MSSM} &\propto& \sin(\phi_{\rm CP})\, \frac{m_l}{M_{\rm
    SUSY}^2}\, \tan\beta\ , 
\end{eqnarray}
where $\phi_{\rm  CP}$ is a  generic soft SUSY-breaking  CP-odd phase.
Nevertheless, beyond the  one-loop approximation~\cite{APEDM,Peskin},
other dependencies of  $d_l$ on $\tan\beta$ are possible  in the MSSM.
However, we show that in  the $\nu_R$MSSM at the one loop level the 
$\tan\beta$ dependence is linear.

\section{Numerical results}
\setcounter{equation}{0}

In  our  numerical  analysis,   we  adopt  the  procedure  established
in~\cite{IPP2013}.   As  a  benchmark  model, we  choose  a  minimally
extended scenario of minimal  supergravity (mSUGRA), in which we allow
for the bilinear and trilinear soft SUSY-breaking terms, ${\bf B}_\nu$
and ${\bf  A}_\nu$, to acquire  at the GUT scale  overall CP-violating
phases denoted as $\theta$  and $\phi$, respectively.  In addition, we
choose the  sign of  the $\mu$-parameter to  be positive.  As  for the
neutrino  Yukawa  coupling  matrix  ${\bf  h}_\nu$,  we  consider  the
approximate    U(1)-    and    $A_4$-symmetric    models    introduced
in~\cite{APRLtau} and~\cite{KS}, respectively. In these two scenarios,
${\bf h}_\nu$  can be expressed in  terms of the  real parameters $a$,
$b$ and $c$ and CP-odd phases that might be relevant for leptogenesis.
Explicitly, the  neutrino Yukawa-coupling matrix ${\bf  h}_\nu$ in the
U(1)-symmetric model is given by~\cite{APRLtau}
\begin{eqnarray}
  \label{YU1}
{\bf h}_\nu &=&
 \left(\begin{array}{lll}
 0 & 0 & 0 \\
 a e^{-\frac{i\pi}{4}} & b e^{-\frac{i\pi}{4}} & c e^{-\frac{i\pi}{4}} \\
 a e^{ \frac{i\pi}{4}} & b e^{ \frac{i\pi}{4}} & c e^{ \frac{i\pi}{4}}
 \end{array}\right)\; ,
\end{eqnarray}
and in the model based on the $A_4$ discrete symmetry by~\cite{KS}
\begin{eqnarray}
  \label{YA4}
{\bf h}_\nu &=&
 \left(\begin{array}{lll}
 a & b & c \\
 a e^{-\frac{2\pi i}{3}} & b e^{-\frac{2\pi i}{3}} & c e^{-\frac{2\pi i}{3}} \\
 a e^{ \frac{2\pi i}{3}} & b e^{ \frac{2\pi i}{3}} & c e^{ \frac{2\pi i}{3}}
\end{array}\right)\; .
\end{eqnarray} 
It should  be noted that the  choices of the  neutrino Yukawa matrices
(\ref{YU1}) and (\ref{YA4}) both  lead to massless light neutrinos for
any  value  of the  heavy  neutrino mass  scale  $m_N$,  as these  are
protected  by the  $U(1)$ and  $A_4$ symmetries.   The  observed light
neutrino  masses and  mixings  can be  obtained  by introducing  small
symmetry-breaking parameters~$\delta_{ij}$ (with $i, j = 1,2,3$), such
that $\delta_{ij} \ll a,\ b,\  c$.  Since lepton dipole moments remain
practically unaffected by these small symmetry-breaking parameters, we
do not consider them here in detail.

For definiteness, our  numerical analysis in this section  is based on
the following baseline scenario:
\begin{eqnarray}
  \label{baseline}
m_0 & =& 1~{\rm TeV}\; ,\qquad 
M_{1/2}\ =\ 1~{\rm TeV}\; , \qquad A_0\ =\ -4~{\rm TeV}\; ,\qquad
\tan\beta \ =\ 20\;,\nonumber\\
m_N & = & 1~{\rm TeV}\;, \qquad B_0\ =\ 0.1~{\rm TeV}\;, \qquad
a\ =\ b\ =\ c\ =\ 0.05\;, 
\end{eqnarray}
where $m_0$, $M_{1/2}$  and $A_0$  are  the standard  universal soft  
SUSY-breaking parameters. All mass parameters except $m_N$ are defined 
at the GUT scale and $m_N$ is intaken at $m_N$ scale. 
It is  understood that  those parameters  not explicitly
quoted   in   the   text    assume   their   default   values   stated
in~(\ref{baseline}).   Likewise,   unless  it  is   explicitly  stated
otherwise, our  default scenario  for ${\bf h}_\nu$  is the  one given
in~(\ref{YA4}), with the  specific choice $a = b = c  = 0.05$ as given
in~(\ref{baseline}).

In the following the $\nu_R$MSSM contributions to the  muon  MDM  not 
present in the SM are denoted by $a_\mu$. 
We  investigate the  dependence of  $a_\mu$ and  $d_e$ on  several key
theoretical parameters,  by varying  them around their  baseline value
given  in~(\ref{baseline}),  while  keeping the  remaining  parameters
fixed. In  doing so, we also  make sure that  the displayed parameters
can  accommodate the  LHC data  for a  SM-like Higgs  boson  with mass
$m_H=125.5\pm 2$ GeV \cite{Hmass} and satisfy the current lower limits
on  gluino  and   squark  masses  \cite{m-gtqt},  i.e.~$m_{\tilde{g}}>
1500$~GeV and  $m_{\tilde{t}}>500$~GeV.  In the  following, we present
numerical results first for $a_\mu$ and then for $d_e$.

\subsection{Results for $a_\mu$}
 
Our  numerical  estimates  for  $a_\mu$  exhibit  a  direct  quadratic
dependence on  the muon mass  $m_\mu$. In fact,  we find that  for the
same set of soft  SUSY-breaking parameters $m_0$, $M_{1/2}$ and $A_0$,
the  ratio  $a_\mu/a_e$  remains  constant to  a  good  approximation,
i.e.~$a_\mu/a_e \approx m^2_\mu /m^2_e  \approx 42752.0$.  In order to
understand this parameter dependence, we have to carefully analyze the
soft SUSY-breaking contributions to the form-factors:
\begin{eqnarray}
  \label{GllgLSB}
G_{ll\gamma}^{L,{\rm SB}}
 &=&
 \tilde{V}^{0 \ell R}_{l m a} \tilde{V}^{0 \ell R*}_{l m a}
 \bigg[
   m_{l} \lambda_{\tilde{e}_a}
   J^1_{41}(\lambda_{\tilde{e}_a},\lambda_{\tilde{\chi}^0_m})
 \bigg]
 +
 \tilde{V}^{0 \ell L}_{l m a} \tilde{V}^{0 \ell L*}_{l m a}
 \bigg[
   m_{l} \lambda_{\tilde{e}_a} J^1_{41}(\lambda_{\tilde{e}_a},\lambda_{\tilde{\chi}^0_m})
 \bigg] 
\nonumber\\
 &&+\
 \tilde{V}^{0 \ell L}_{l m a} \tilde{V}^{0 \ell R*}_{l m a}
 \bigg[
   2 m_{\tilde{\chi}^0_m} \lambda_{\tilde{e}_a}
   J^0_{31}(\lambda_{\tilde{e}_a},\lambda_{\tilde{\chi}^0_m})
 \bigg]\; ,\\
    \label{GllgRSB}
G_{ll\gamma}^{R,{\rm SB}}
 &=&
 \tilde{V}^{0 \ell L}_{l m a} \tilde{V}^{0 \ell L*}_{l m a}
 \bigg[
   m_{l} \lambda_{\tilde{e}_a}
   J^1_{41}(\lambda_{\tilde{e}_a},\lambda_{\tilde{\chi}^0_m})
 \bigg]
 +
 \tilde{V}^{0 \ell R}_{l m a} \tilde{V}^{0 \ell R*}_{l m a}
 \bigg[
   m_{l} \lambda_{\tilde{e}_a} J^1_{41}(\lambda_{\tilde{e}_a},\lambda_{\tilde{\chi}^0_m})
 \bigg]\nonumber\\
 &&+\ 
 \tilde{V}^{0 \ell R}_{l m a} \tilde{V}^{0 \ell L*}_{l m a}
 \bigg[
   2 m_{\tilde{\chi}^0_m} \lambda_{\tilde{e}_a}
   J^0_{31}(\lambda_{\tilde{e}_a},\lambda_{\tilde{\chi}^0_m})
 \bigg]\; ,
\end{eqnarray}
where   the  different   terms  that   occur   in~(\ref{GllgLSB})  and
(\ref{GllgRSB}) are defined  in~\cite{IPP2013} and are also explicitly
given  in~Appendix~\ref{f3.4and3.5}.    Observe  that  the  neutralino
vertices induce  a term  which is not  manifestly proportional  to the
charged lepton  mass, but to  the neutralino mass.  However,  a closer
inspection of  the products of the mixing  matrices $\tilde{V}^{0 \ell
  R}_{l m  a} \tilde{V}^{0  \ell R*}_{l m  a}$ and  $\tilde{V}^{0 \ell
  R}_{l m a} \tilde{V}^{0  \ell L*}_{l m a}$ reveals~\cite{Sto06} that
these last  expressions are by themselves proportional  to the charged
lepton mass  $m_l$.  The latter provides a  non-trivial powerful check
for the correctness of the results presented here.

\begin{figure}[!ht]
 \centering
 \includegraphics[clip,width=0.30\textwidth]{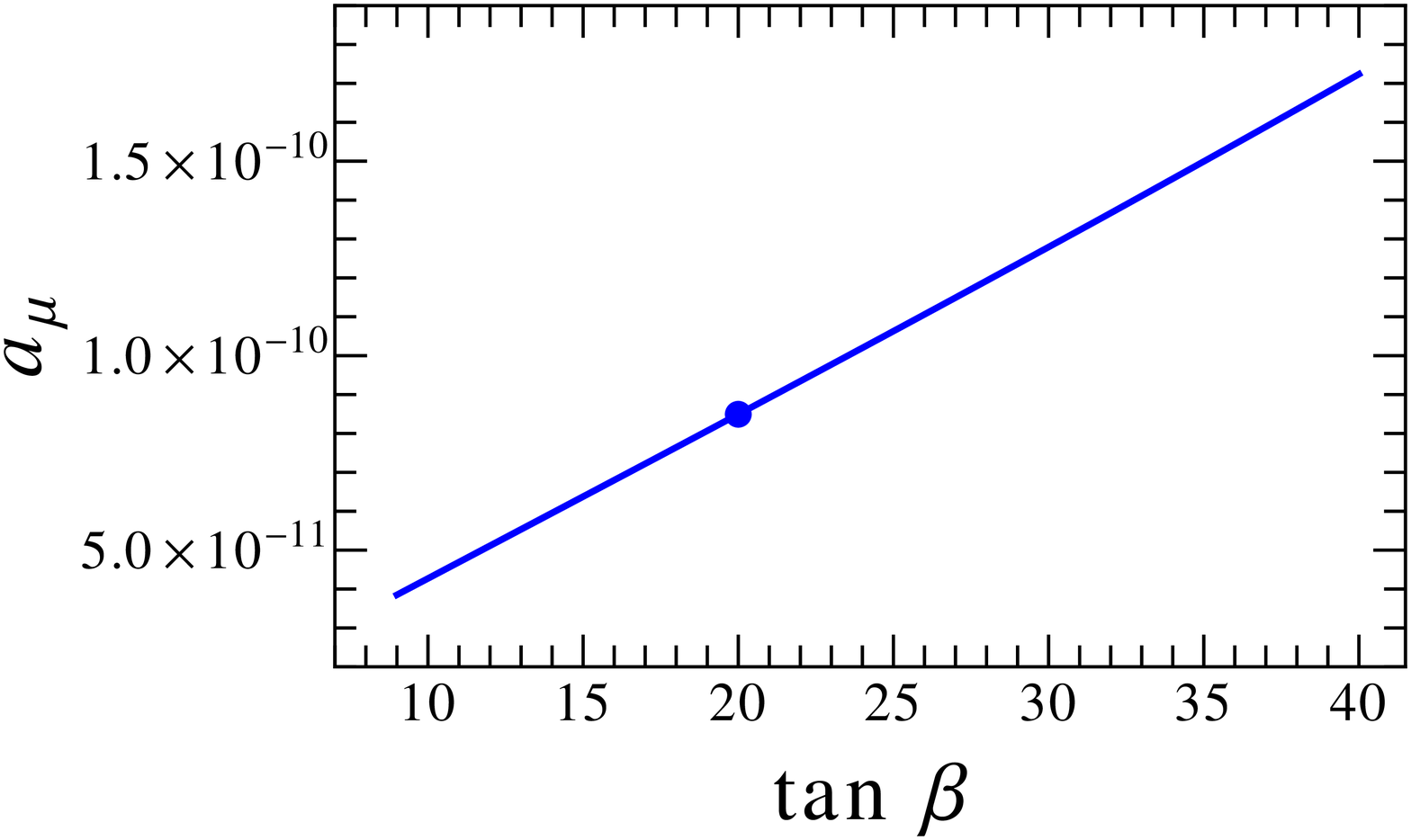} \hspace{1cm}
 \includegraphics[clip,width=0.30\textwidth]{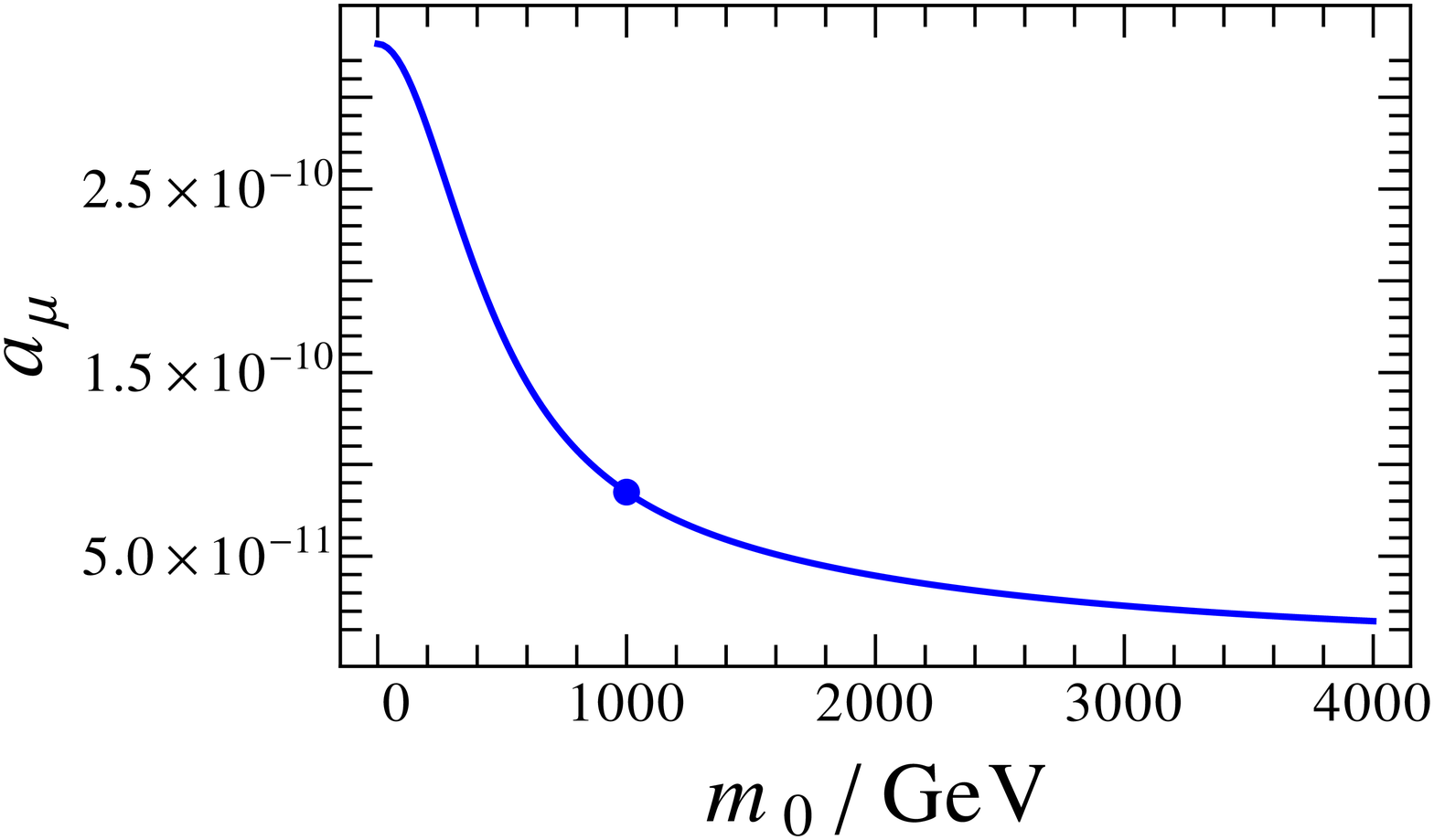}
 \\
 {\bf \footnotesize \hspace{3em} (a) \hspace*{17em} (b)}
 \\[2ex]
 \includegraphics[clip,width=0.30\textwidth]{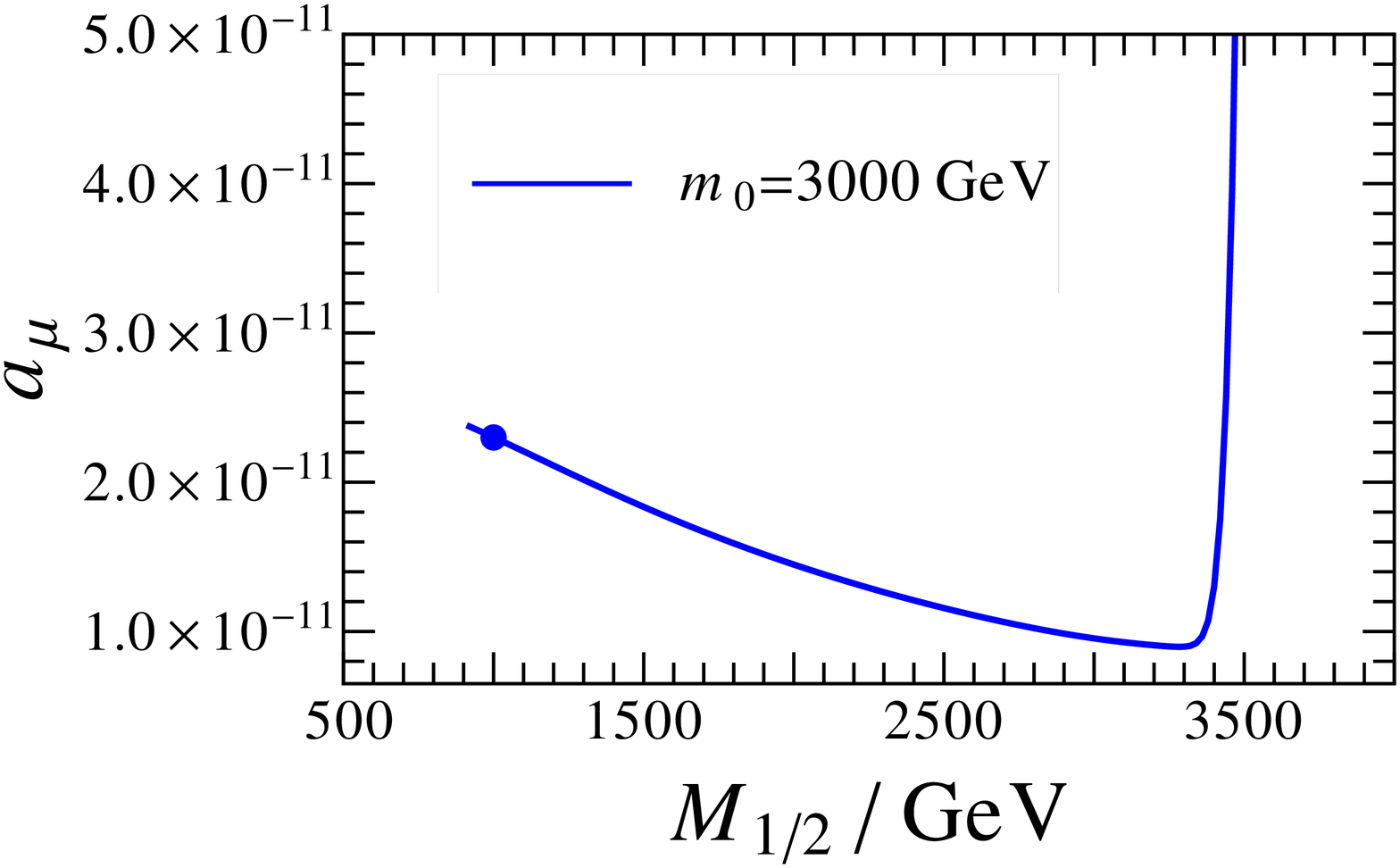} \hspace{1cm}
 \includegraphics[clip,width=0.30\textwidth]{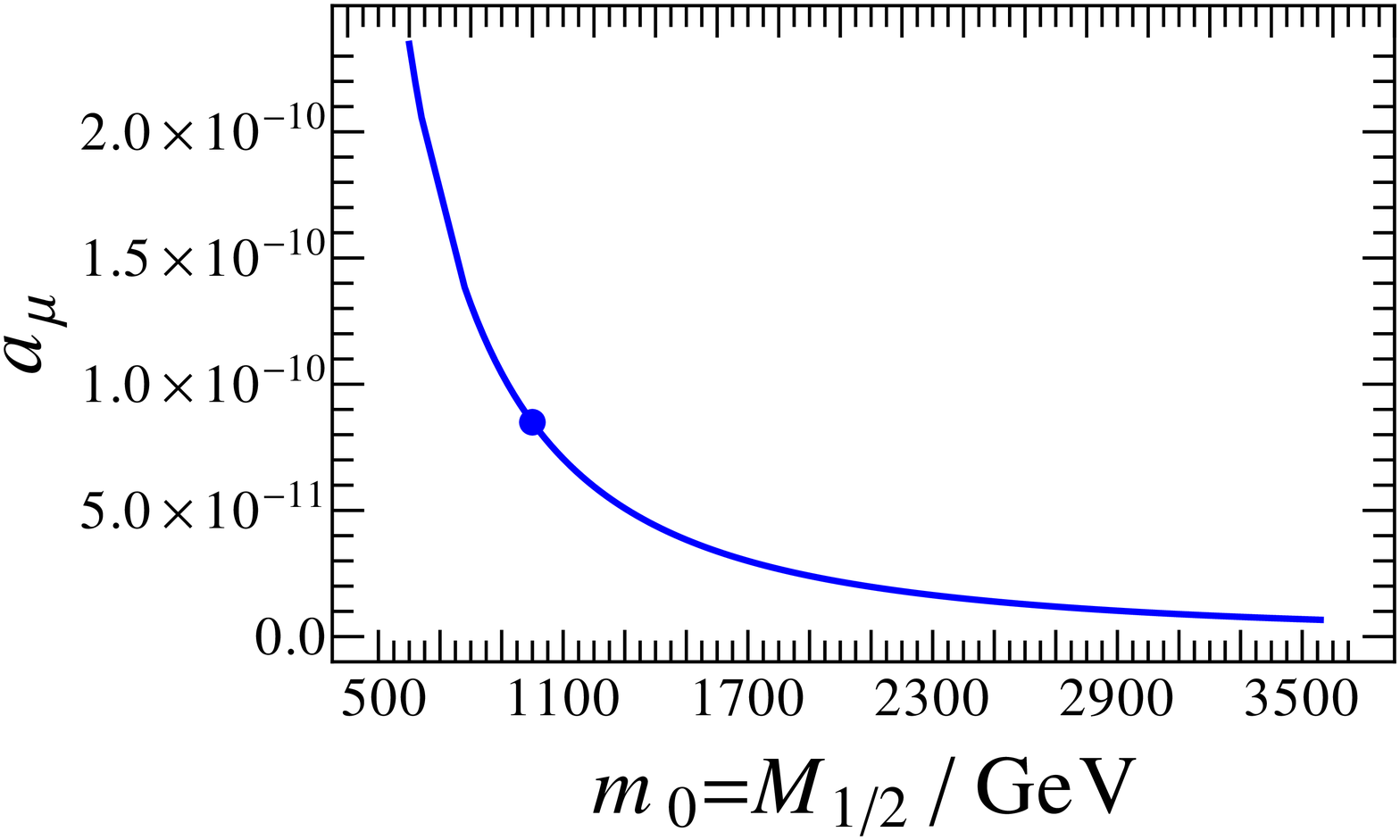}
 \\
 {\bf \footnotesize \hspace{3em} (c) \hspace*{17em} (d)}
 \\[2ex]
 \includegraphics[clip,width=0.30\textwidth]{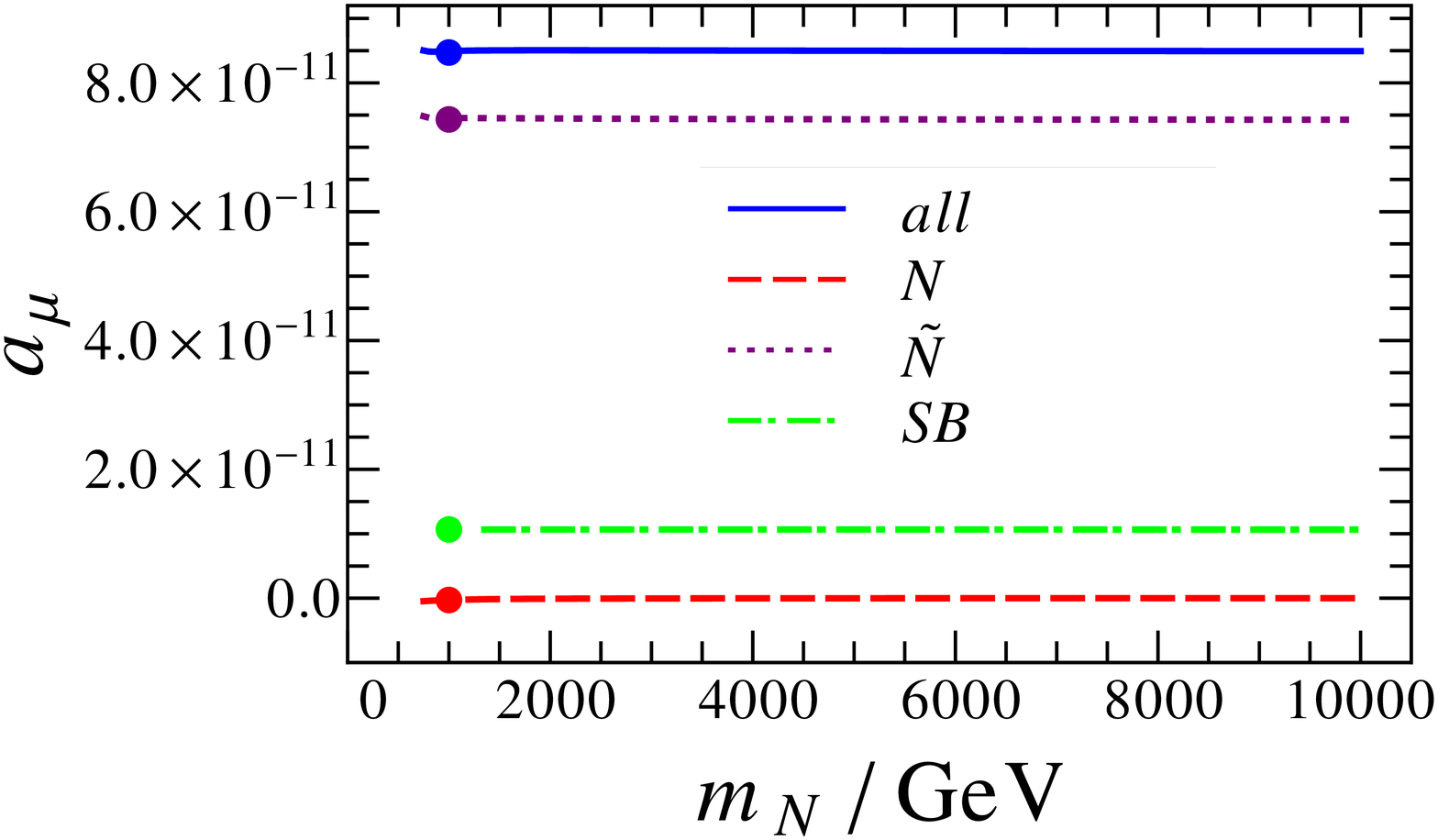} \hspace{1cm}
 \includegraphics[clip,width=0.30\textwidth]{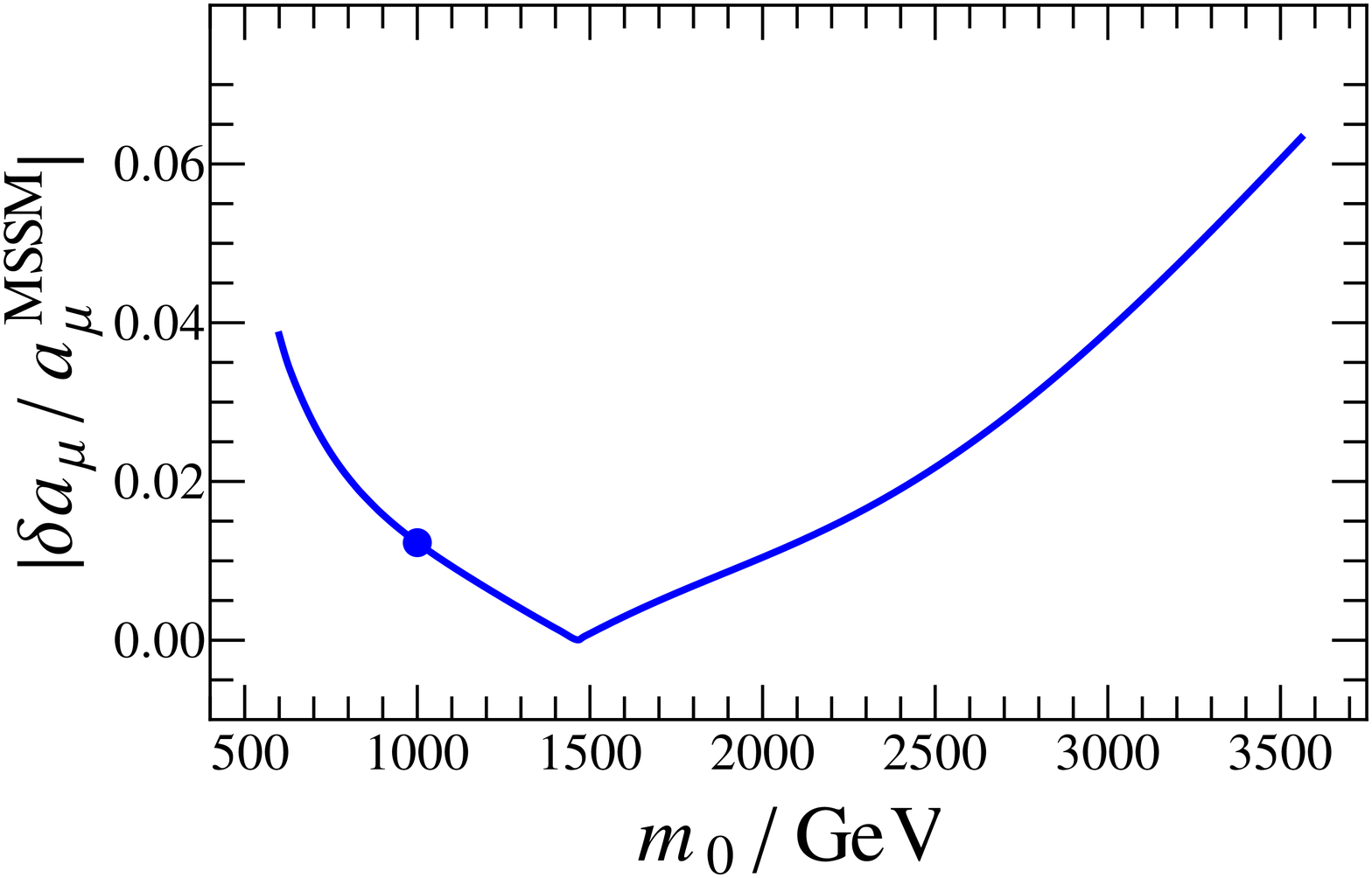}
 \\
 {\bf \footnotesize \hspace{3em} (e) \hspace*{17em} (f)}
 \\[2ex]
\caption{Numerical  estimates for the  muon anomalous  MDM~$a_\mu$, as
  functions  of  $\tan\beta$,  $M_{1/2}$,  $m_N$,  $m_0$  and  $m_0  =
  M_{1/2}$,  in the  $\nu_R$MSSM.  The  default parameter  set  of the
  baseline model  is given in~(\ref{baseline}).  The  pannel (e) shows
  the    heavy     neutrino~($N$),    sneutrino~($\tilde{N}$),    soft
  SUSY-breaking~(SB)  and {\it  all}  contributions to  $a_\mu$, as  a
  function of  $m_N$. The pannel  (f) displays an absolute value of the  relative deviation
  $\delta  a_\mu/a_\mu$ of  the $\nu_R$MSSM  and MSSM  predictions for
  $a_\mu$ [cf.~(\ref{damu})],  as a function  of $m_0$.  The  range of
  input parameters in all plots satisfy the current LHC constraints on
  Higgs, gluino and squark masses.   The heavy dots on the curves give
  the  predicted   values  for  $a_\mu$  evaluated   for  the  default
  parameters (\ref{baseline}).}
\label{Fig1}
\end{figure}

In addition, our numerical analysis  shows that the muon anomalous MDM
$a_\mu$ is  almost independent of the  neutrino-Yukawa parameters $a$,
$b$  and $c$, the  heavy neutrino  mass $m_N$  and the  soft trilinear
parameter  $A_0$.  Hence,  our  results are  almost  insensitive to  a
particular  choice  for  a  neutrino  Yukawa  texture,  e.g.~as  given
in~(\ref{YU1})  and~(\ref{YA4}), and  also independent  of  the CP-odd
phases $\theta$ and $\phi$.
 
In  Fig.~\ref{Fig1},  we  give  numerical estimates  for  $a_\mu$,  as
functions of  the key theoretical  parameters: $\tan\beta$, $M_{1/2}$,
$m_0$ and~$m_N$.  In the frame (a) of this figure, we see that $a_\mu$
depends linearly  on $\tan\beta$, as expected from~(\ref{al_approx}). 
Likewise, we  have investigated  in Fig.~\ref{Fig1} the  dependence of
$a_\mu$ on the soft  SUSY-breaking parameters $m_0$ and $M_{1/2}$, for
different kinematic  situations, and obtained  results consistent with
the scaling behaviour of~$1/M^2_{\rm SUSY}$ in~(\ref{al_approx}).

In the  pannel (e) of Fig.~\ref{Fig1},  we observe that  the effect of
the heavy right-handed neutrinos~($N$) and sneutrinos~($\tilde{N}$) on
$a_\mu$  is negative,  but  small, in  agreement  with our  discussion
above.  The size  of their contributions alone to  $a_\mu$ ranges from
$-10^{-12}$ to  $-4.8\times 10^{-15}$, for  $m_N = 0.5$ -  10~TeV.  On
the other hand, the left-handed sneutrino contributions to $a_\mu$ are
approximately independent  of the heavy Majorana  mass $m_N$, reaching
values   $\approx  8.5\times   10^{-11}$.    The  soft   SUSY-breaking
contributions are also approximately independent of the heavy Majorana
mass~$m_N$ and  have values  $\approx 1.1\times 10^{-12}$.   Note that
the light  sneutrino contribution to the anomalous  magnetic moment is
the  largest in  magnitude,  and it  is  already present  in the  MSSM
contributions to  $a_\mu$.  
Finally, we have checked  the dominance of the MSSM contributions by 
looking at the dependence of the parameter:
\begin{equation}
  \label{damu}
\delta a_\mu \ =\ a^{\rm \nu_RMSSM}_\mu\ -\ a^{\rm MSSM}_\mu\; .
\end{equation}
The difference  $\delta a_\mu$ of  the predictions for  $a_\mu$ within
the  $\nu_R$MSSM  and the  MSSM  divided by $a_\mu$ is  evaluated,  
and the absolute values of the results  are
displayed  in the  pannel (f)  of  Fig.~\ref{Fig1}, as  a function  of
$m_0=M_{1/2}$.  The largest deviation from the MSSM is found for 
largest allowed parameter value, $m_0=3600$~GeV, in which case
$\delta a_\mu/a^{\rm MSSM}_\mu$ is as large as $6.2\times 10^{-2}$.

\subsection{Results for $d_e$}

We now study  the dependence of the electron EDM  $d_e$ on several key
model parameters, such as $m_0$, $M_{1/2}$, $B_0$, $A_0$, $\tan\beta$,
$\theta$ and  $\phi$. The predictions  for $d_\mu$ may be  obtained by
using  the naive  scaling relation:  $d_\mu \approx  (m_\mu/m_e)\, d_e
\approx 205\, d_e$.  We  have found this scaling behaviour is 
numerically satisfied very well. The maximal numerical values for 
$d_e$ we obtained are of the order $\sim 10^{-27}$ e$\;$cm. Therefore 
predicted values  for $d_\mu$ are  always found to be  less than~$\sim
10^{-25}\ e\;  {\rm cm}$, which  is several orders of  magnitude below
the  present  experimental   upper  bound:  $d_\mu=  0.1\pm  0.9\times
10^{-19}\ e\; {\rm cm}$ \cite{PDG2013}.

We note that  heavy singlet neutrinos $N$ do  not contribute to $d_e$,
even if the  soft SUSY-breaking CP-odd phases $\phi$  and $\theta$ are
non-zero.  On the  other hand, soft SUSY-breaking and 
right handed neutrino effects induce non-vanishing  $d_e$,  if  either  
$\theta$ or  $\phi$  are  non-zero. If both $\phi=0$ and $\theta=0$, 
lepton EDMs $d_l$ numerically vanish. Therefore, the complex products of vertices 
(\ref{CPNt}) emerging in the $\nu_R$MSSM do not induce the CP 
violation at one loop level, in accord with the result of Ref. 
\cite{Peskin} obtained in the MSSM with a high-scale seesaw mechanism.
\begin{figure}[!ht]
 \centering
 \includegraphics[clip,width=0.30\textwidth]{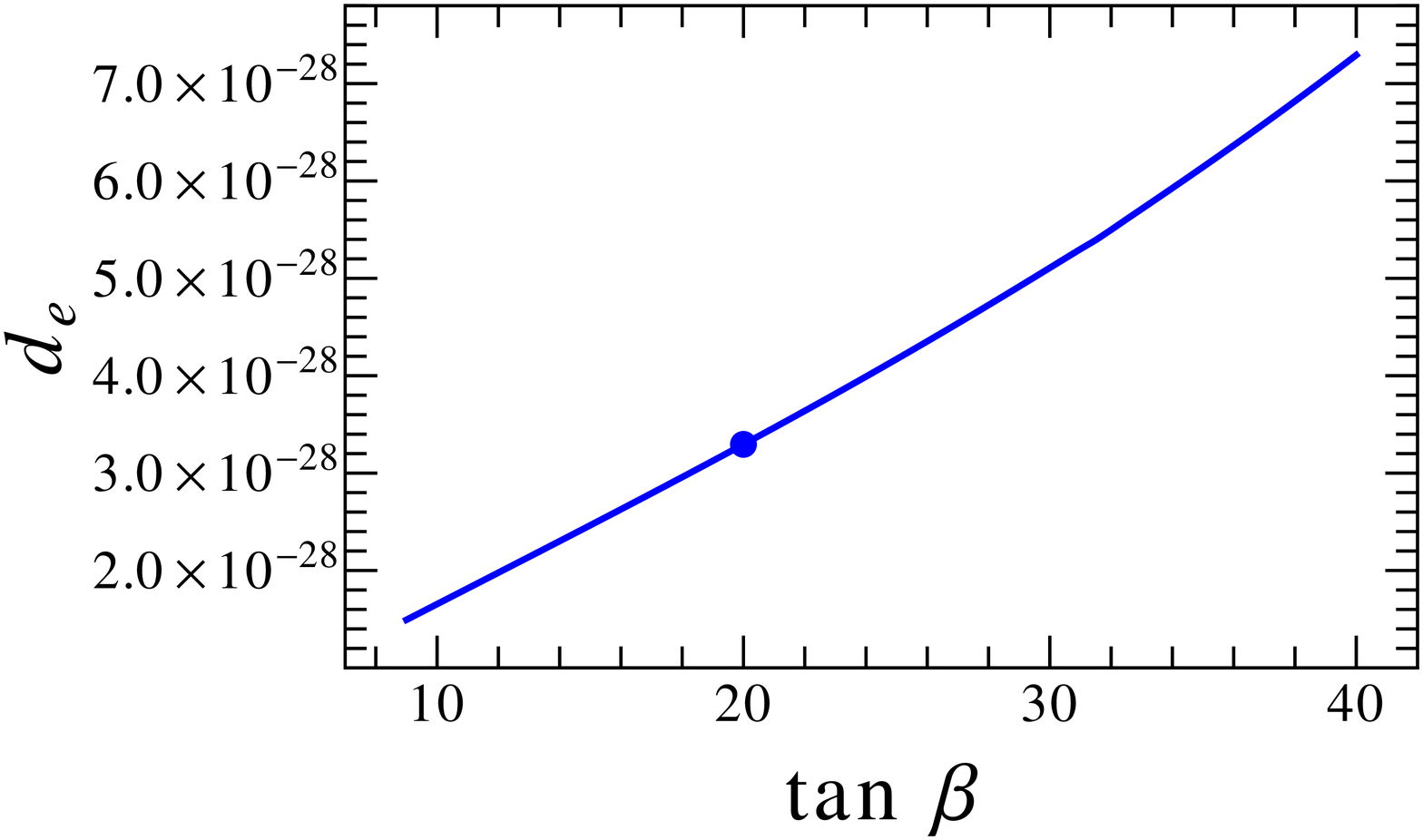} \hspace{1cm}
 \includegraphics[clip,width=0.30\textwidth]{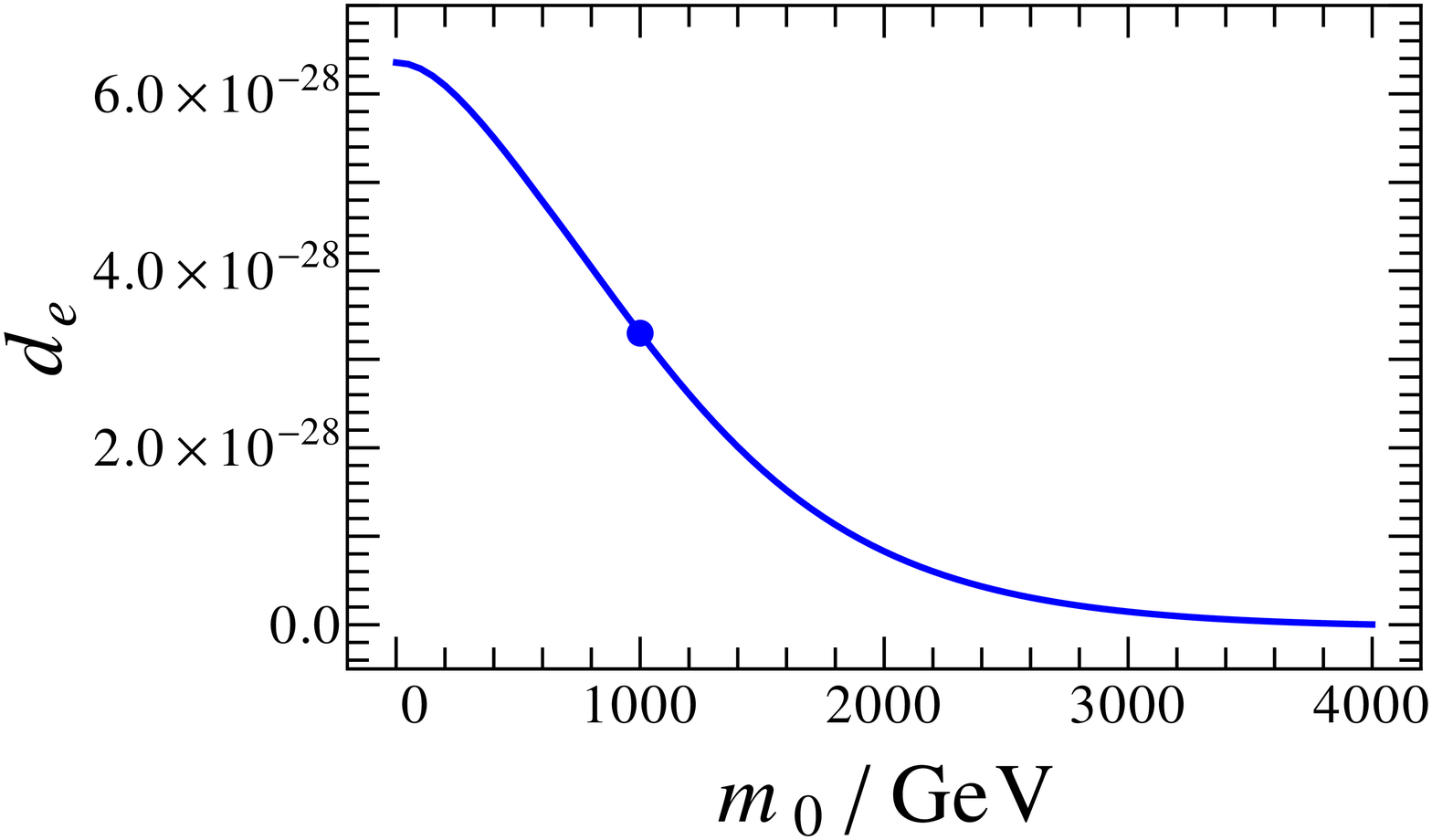}
 \\
 {\bf \footnotesize \hspace{3em} (a) \hspace*{17em} (b)}
 \\[2ex]
 \includegraphics[clip,width=0.30\textwidth]{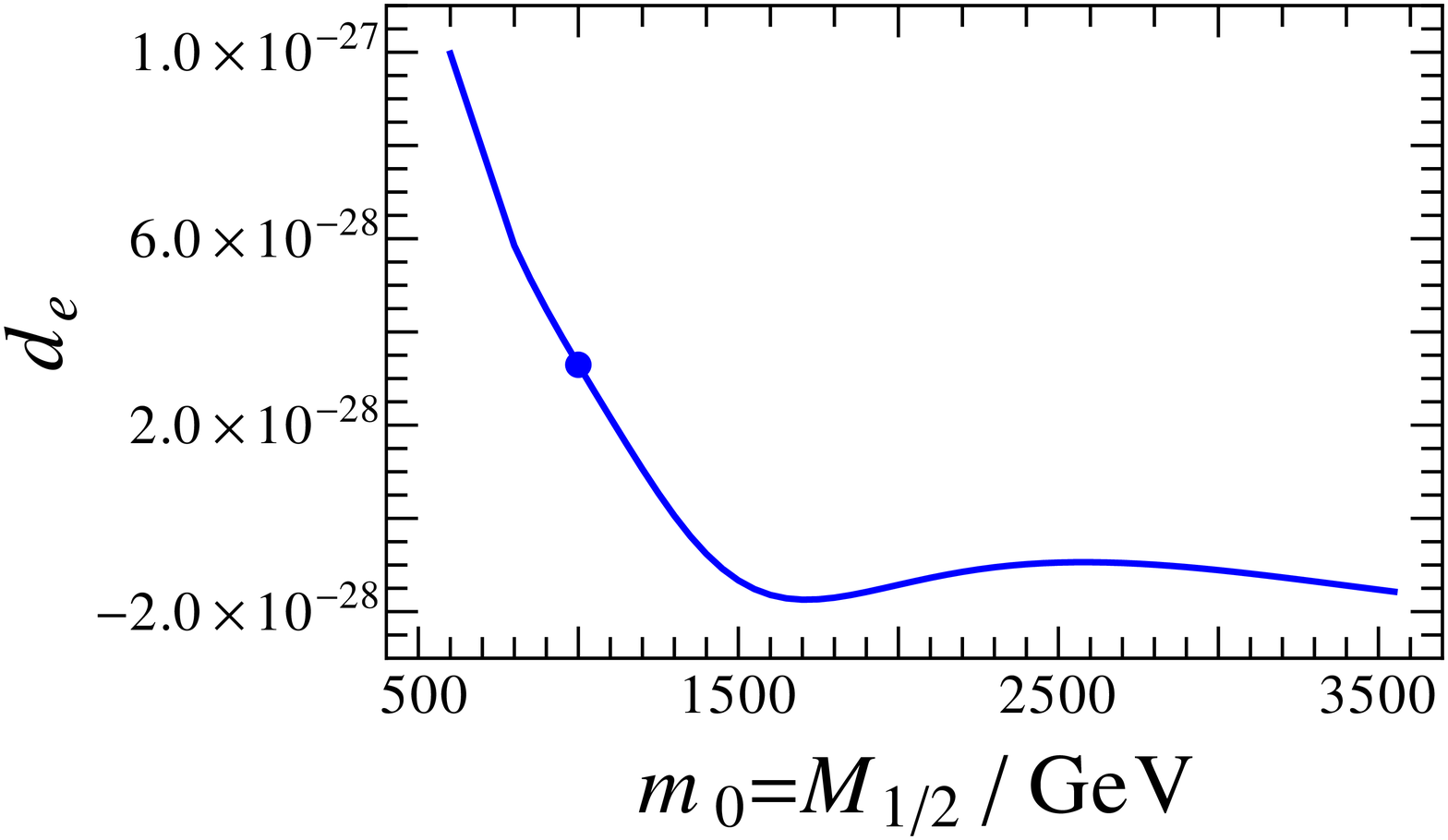} \hspace{1cm}
 \includegraphics[clip,width=0.30\textwidth]{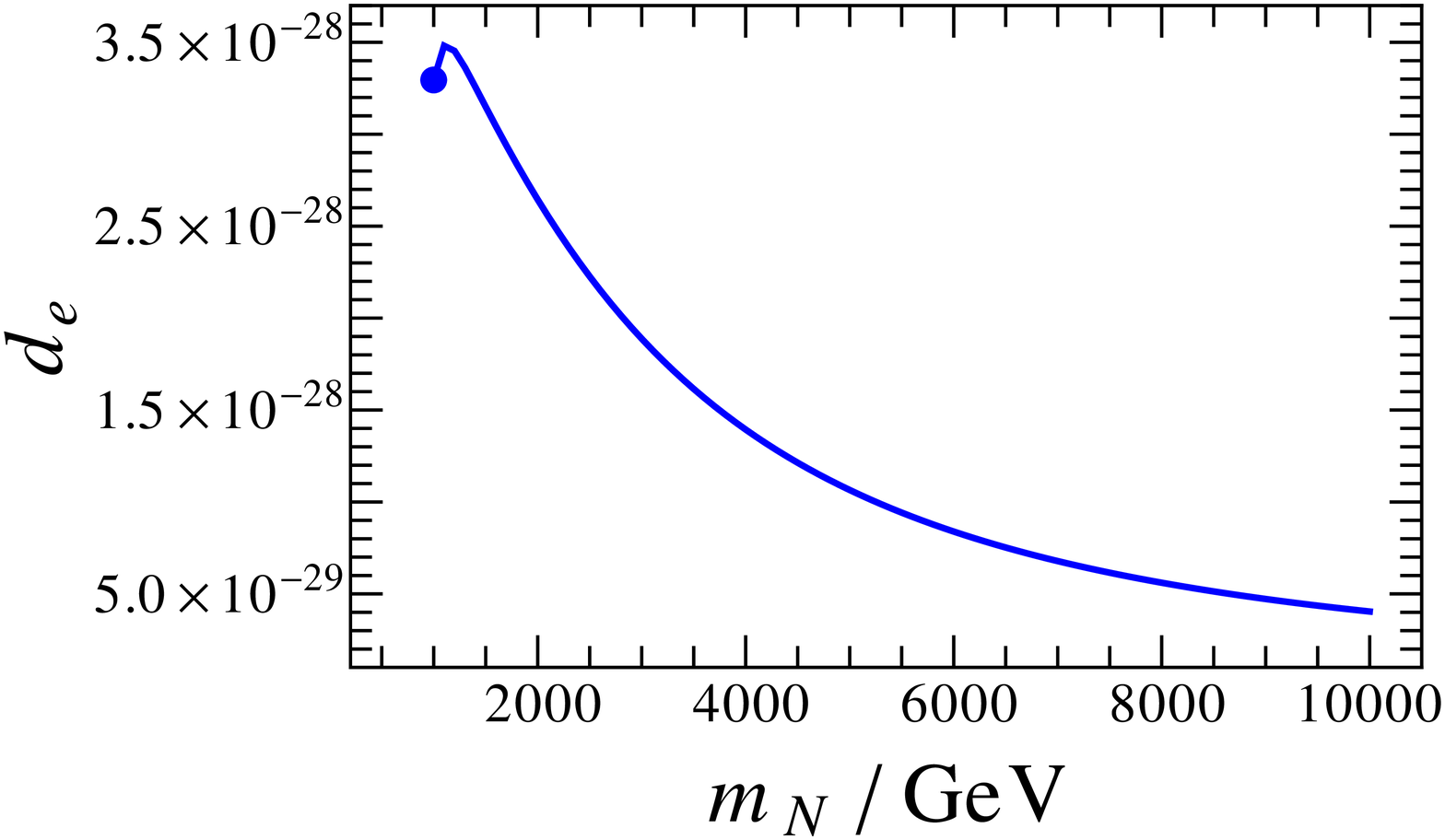}
 \\
 {\bf \footnotesize \hspace{3em} (c) \hspace*{17em} (d)}
 \\[2ex]
\caption{Numerical  estimates   of  the  electron   EDM~$d_e$  in  the
  $\nu_R$MSSM, as  functions of $\tan\beta$,  $m_0$, $m_0=M_{1/2}$ and
  $m_N$, for $\phi=\pi/2$. The remaining parameters not 
  shown assume the baseline values in~(\ref{baseline}). All
  input parameters are chosen so  as to satisfy the LHC constraints on
  Higgs,  gluino and  squark masses.   The  heavy dots  on the  curves
  indicate the  predicted values for  $d_e$ evaluated for  the default
  parameters (\ref{baseline}).}
\label{Fig2}
\end{figure}

In  Fig.~\ref{Fig2}, we present  numerical estimates  of $d_e$  on the
$\nu_R$MSSM  parameters $\tan\beta$, $m_0$,  $M_{1/2}$ and  $m_N$, for
maximal $A_0$ phase, $\phi = \pi/2$.   We also set $\theta =  0$, since the
dependence of $d_e$  on $B_0$ is weaker than  the dependence on $A_0$.
As shown in pannel (a) Fig $d_e$ exhibits a linear 
dependence on $\tan\beta$ confirming the $\tan\beta$ naive scaling 
behaviour in Eq. (\ref{di_approx}).
Further, $d_e$ is a decreasing function of $m_0$. As a function of 
$m_0=M_{1/2}$, $d_e$ assumes both positive and negative values, and is roughly
proportional to $-1 -2.4\,\mathrm{TeV}/m_0 + 6.3\mathrm{TeV}^2/m_0^2$.
There is also  a small region of parameter space
for  $m_0=M_{1/2}   \stackrel{<}{{}_\sim}  800$~GeV,  for   which  the
prediction for $d_e$ is of the order of the  experimental upper limit on 
$d_e$ (\ref{deUB}).
In addition, $d_e$ decreases with increasing $m_N$: for the $m_N$ values 
from the pannel (d) of Fig.~\ref{Fig2} this behavior can
roughly approximated by a function $-0.13+\mathrm{TeV}^{\frac{2}{3}}m_N^{-\frac{2}{3}}$, 
in the $m_N$-range $10\,\mathrm{TeV}<m_N<100\,\mathrm{TeV}$ $d_e$ roughly scales as $1/m_N$, and above
$m_N=100\,\mathrm{TeV}$ it becomes very slowly decreasing function in $m_N$.

\begin{figure}[!t]
 \centering
 \includegraphics[clip,width=0.30\textwidth]{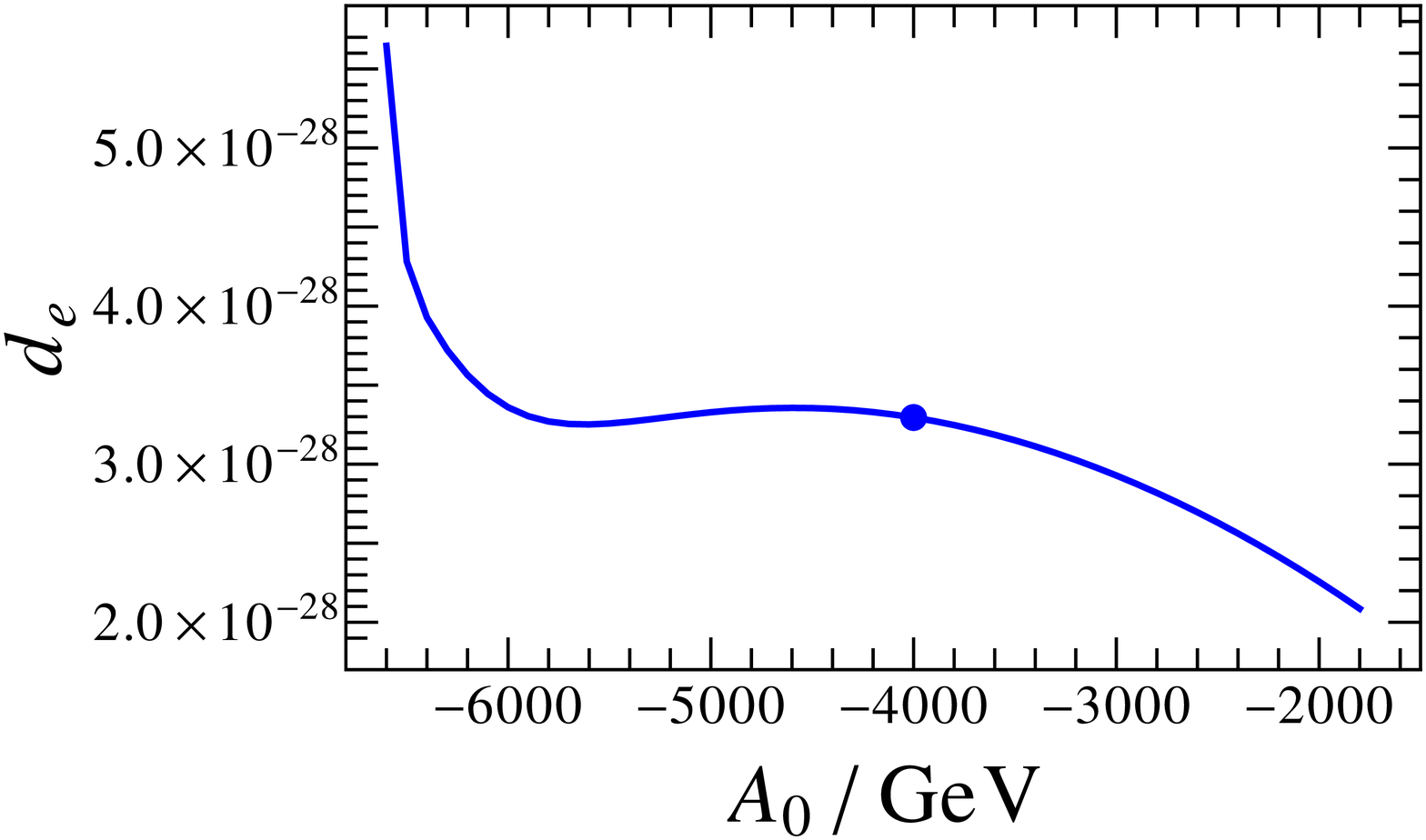}\hspace{1cm}
 \includegraphics[clip,width=0.30\textwidth]{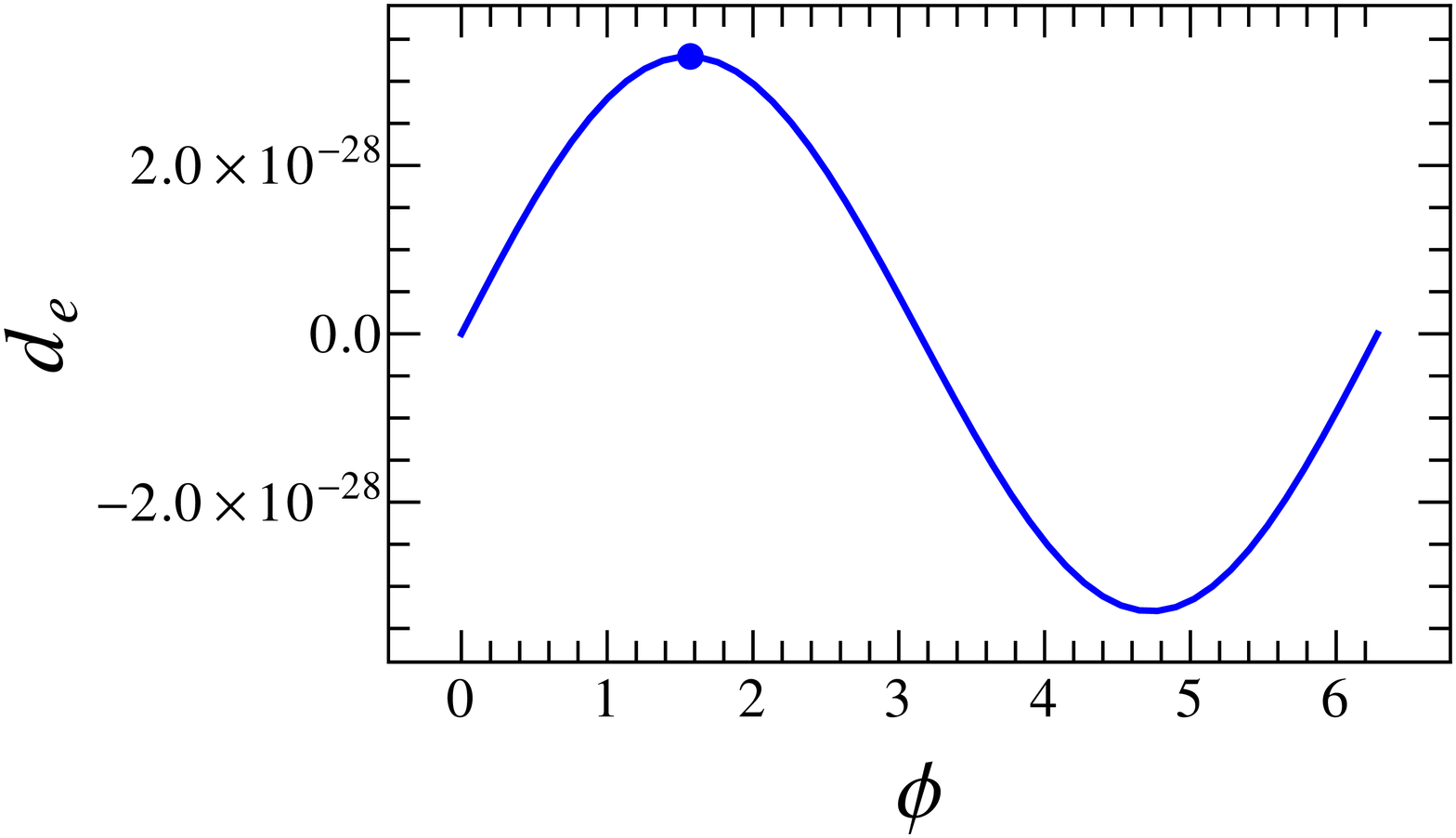} 
 \\
 {\bf \footnotesize \hspace{3em} (a) \hspace*{18em} (b)}
 \\[2ex]
 \includegraphics[clip,width=0.30\textwidth]{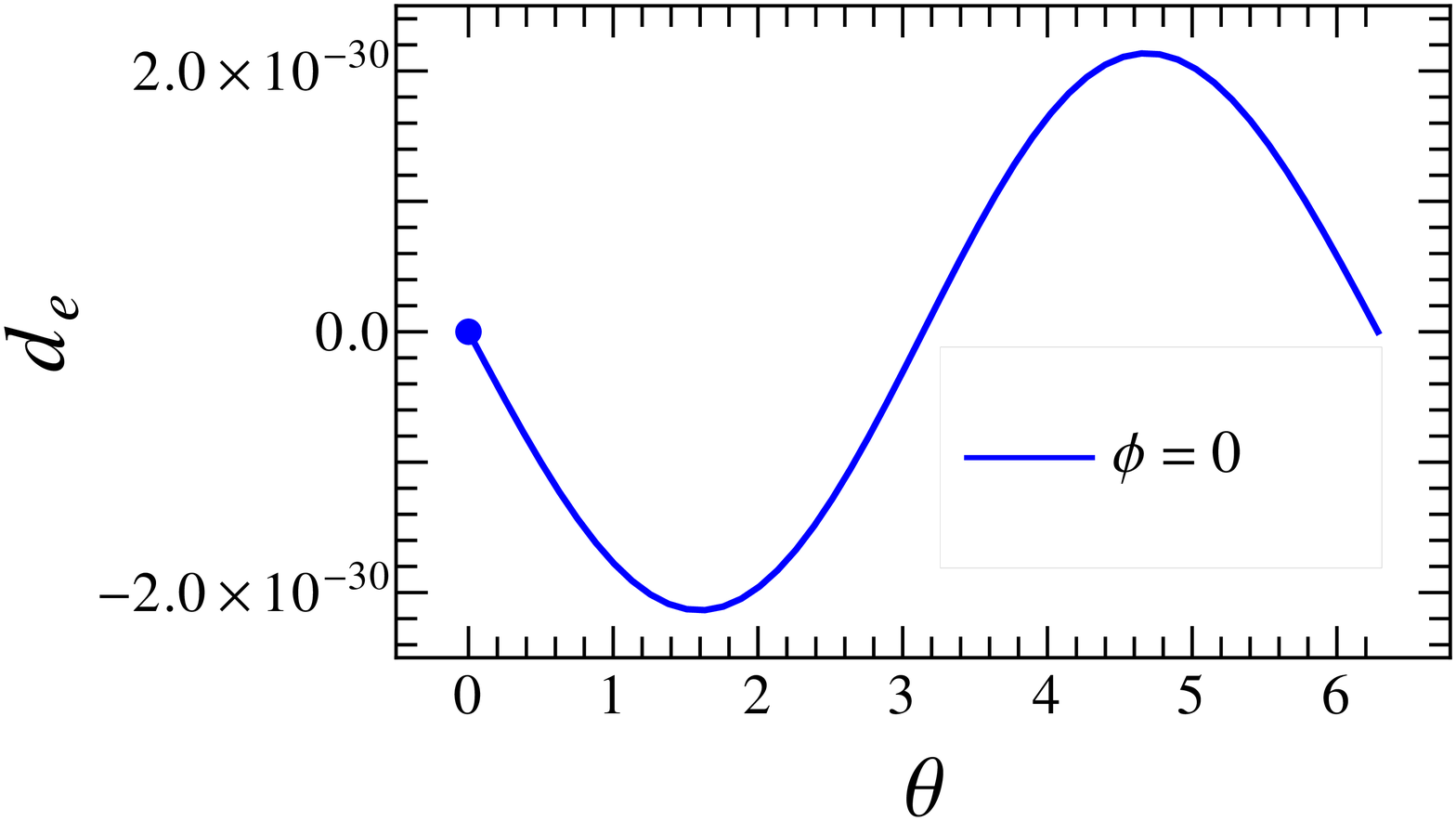}\hspace{1.5cm}
 \includegraphics[clip,width=0.30\textwidth]{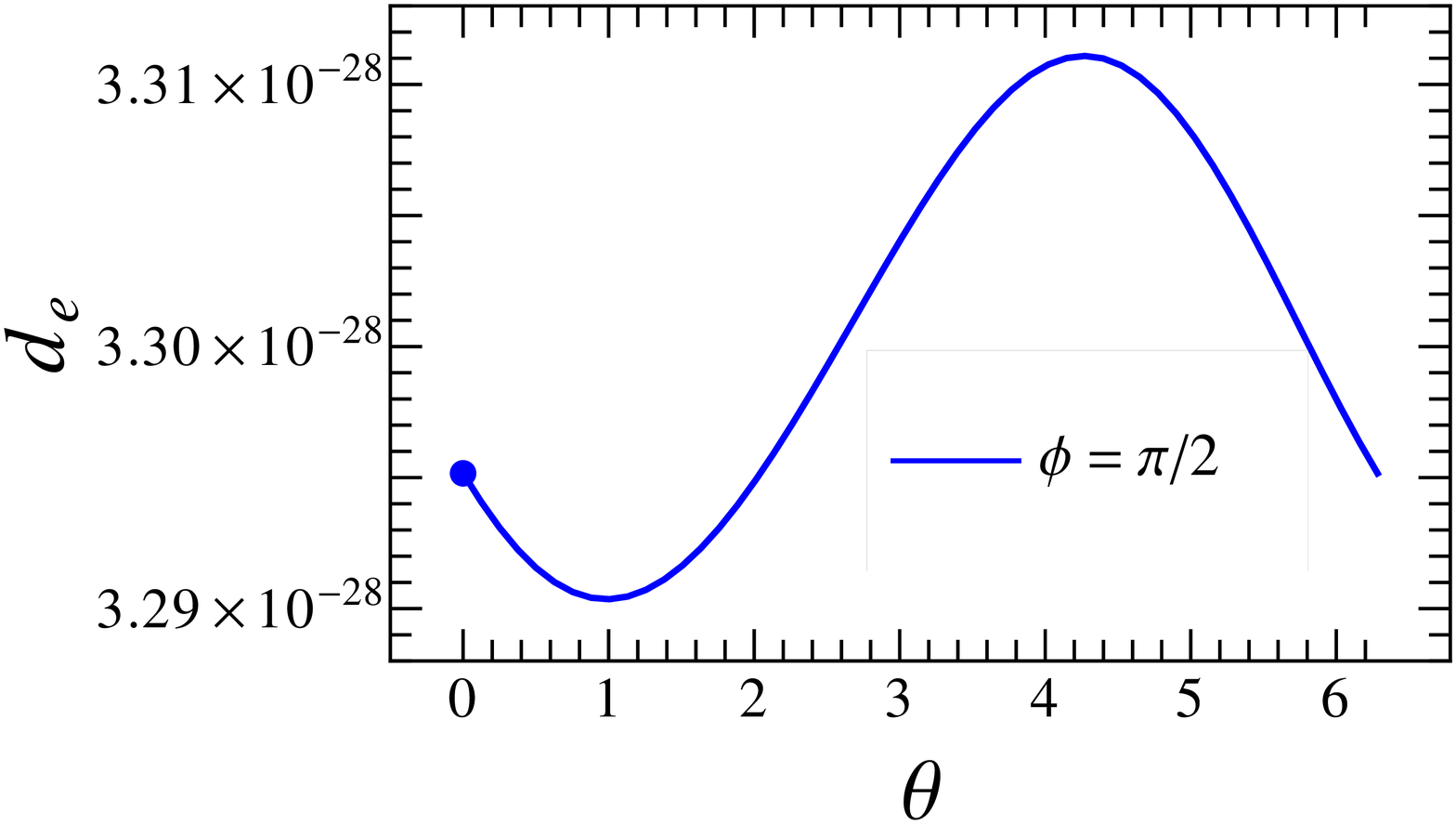}\hspace*{.5cm}
 \\
 {\bf \footnotesize \hspace{3em} (c) \hspace*{18em} (d)}
 \\[2ex]
 \includegraphics[clip,width=0.30\textwidth]{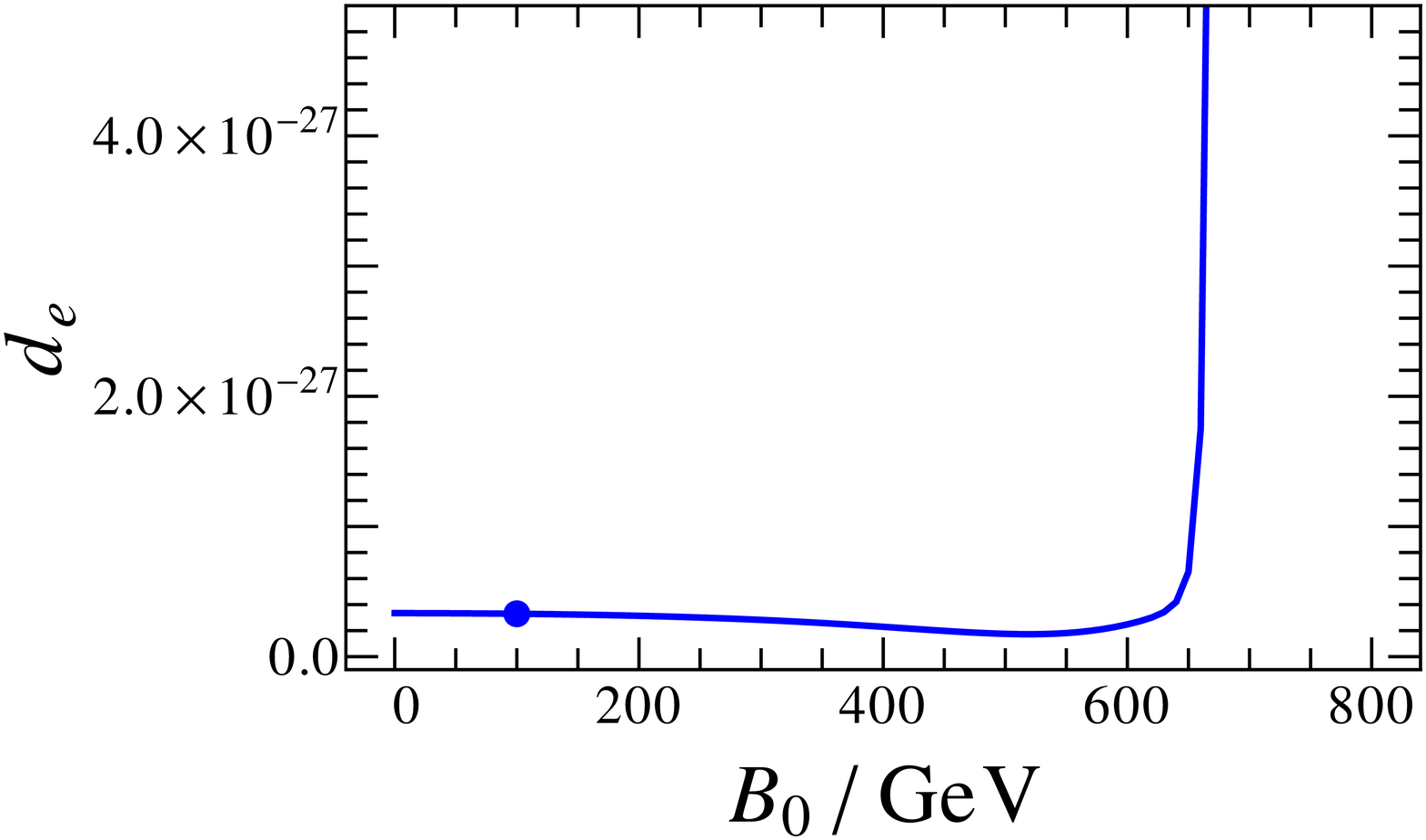}\hspace*{7cm}
 \\
 {\bf \footnotesize \hspace{3em} (e) \hspace*{19em} {}}
 \\[2ex]
\caption{Predicted numerical values  for the electron EDM~$d_e$ versus
  the  soft  SUSY  breaking  parameters  $A_0$  and  $B_0$  and  their
  corresponding  soft  CP-odd  phases   $\phi$  and  $\theta$  in  the
  $\nu_R$MSSM,  for the  baseline  scenario in~(\ref{baseline}). 
  If not shown $\phi$ assumes value $\pi/2$.   The
  range of input parameters shown  in the plots is compatible with the
  LHC constraints on  Higgs, gluino and squark masses.  The heavy dots
  show the  predicted values for  $d_e$, using the  default parameters
  (\ref{baseline}).}
\label{Fig3}
\end{figure}

In Fig.~\ref{Fig3}, we show  the predicted numerical values for $d_e$,
as functions of the soft SUSY-breaking parameters $A_0$ and $B_0$, and
their corresponding CP phases $\phi$  and $\theta$. In all pannels
except the pannel (c), where $\phi=0$ and $\theta$ is a variable, 
$\phi$ assumes value $\pi/2$ or it is a variable
and  $\theta$ is taken to be equal zero. In the pannel (a)
of Fig.~\ref{Fig3}, the soft  trilinear parameter $A_0$ is constrained
by the  LHC data  pertinent to Higgs,  gluino and squark  masses.  
The electron EDM $d_e$ is a complicated function of $|A_0|$ that 
slowly rises for $|A_0|$ between $1.8$ TeV and $4.5$ TeV, slowly 
decreases for $|A_0|$ between $4.5$ TeV and $6$ TeV, and steeply 
rises for $|A_0|>6$ TeV. This function cannot be precisely described 
by a simple Laurent series in $|A_0|$, but in the largest part of the allowed 
$|A_0|$ interval it can roughly be approximated by a constant. The $\phi$
dependence  of $d_e$  is almost  sinusoidal with  an amplitude few times
smaller than the experimental upper bound  (\ref{deUB}). Moreover,
$d_e$  is approximately constant function  of $B_0$, up  to 
$B_0\approx  600$~GeV.  For  larger values, 
i.e.~$B_0\stackrel{>}{{}_\sim} 600$~GeV,  $d_e$ steeply rises, somehow
hinting  at a  numerical  instability in  the  diagonalization of  the
sneutrino mass  matrix, so our results  in this regime  are not valid.
For $\phi=\pi/2$, the  electron EDM $d_e$ attains values  of order the
experimental upper limit~(\ref{deUB}), but  for $\phi=\theta = 0$, the
predictions are numerically consistent with zero.  The dependence of
$d_e$ on $\theta$ is sinusoidal with an amplitude of order few $\times
10^{-30}$,  while its  average value  strongly depends  on  the chosen
value  $\phi$.  From  Figs.~\ref{Fig2} and  \ref{Fig3},  the following
dependence of $d_l$ on $m_l$, $m_0=M_{1/2}$, $m_N$ and $\tan\beta$ may be 
deduced
\begin{equation}
  \label{dlapp1}
d_l\ \propto\ \tan\beta\times m_l\times \frac{f(m_0)}{m_N^x},\qquad m_N<10\,\mathrm{TeV}, 
\end{equation}
where $x$ assumes values between $2/3$ and $1$, and $f(m_0)$ is roughly 
proportional to the function $-1 -2.4\,\mathrm{TeV}/m_0 + 6.3\mathrm{TeV}^2/m_0^2$. 
The last factor in Eq. (\ref{dlapp1}) corresponds to the scaling factor $1/M_{SUSY}^2$ 
in the naive approximation (\ref{di_approx}), and 
in the approximate expressions for lepton EDM derived in~\cite{Peskin}.

\section{Conclusions}

We have systematically studied  the one-loop contributions to the muon
anomalous MDM  $a_\mu$ and the electron EDM~$d_e$  in the $\nu_R$MSSM.
In particular,  we have  paid special attention  to the effect  of the
sneutrino  soft  SUSY-breaking  parameters,  ${\bf B_\nu}$  and  ${\bf
  A_\nu}$,  and their  universal CP  phases, $\theta$  and  $\phi$, on
$a_\mu$ and $d_e$. To the best of our knowledge, lepton dipole moments
have not been  analyzed in detail before, within  SUSY models with low
scale singlet (s)neutrinos.

For the  anomalous MDM  $a_\mu$ of  the muon, we  have found  that the
heavy  singlet neutrino  and  sneutrino contributions  to $a_\mu$  are
small, typically one to two orders of magnitude below the muon anomaly
$\Delta a_\mu$.  Instead, left-handed  sneutrinos and sleptons give the
largest effect on $\Delta a_\mu$, exactly  as is the case in the MSSM.
The dependence of  $a_\mu$ on the muon mass~$m_\mu$, $\tan\beta$ and
the soft  SUSY-breaking mass scale $M_{\rm SUSY}$  have been carefully
analyzed  and their  scaling behaviour  according to~(\ref{al_approx})
has  been  confirmed.  Finally,  the  dependence  of  $a_\mu$ on  the
universal   soft  trilinear   parameter~$A_0$,  the   neutrino  Yukawa
couplings  ${\bf  h}_\nu$  and  the  heavy  neutrino  mass  $m_N$  are
negligible.

Furthermore,  we   have  analyzed  the  electron  EDM   $d_e$  in  the
$\nu_R$MSSM. The  heavy singlet neutrinos do not  contribute to $d_e$,
and soft SUSY-breaking and sneutrino terms contribute  only, if the 
phases $\phi$ and/or $\theta$  have a nonzero  value.
The contribution from the possible CP violating 
terms arising from the relatively complex products of the vertices 
exposed in (\ref{CPNt}) is numerically shown
to be equal zero. 
On  the  other  hand,  the
contribution due to a non-zero value  of $\phi$ is the largest and may
give  rise to  values for  the electron  EDM $d_e$  comparable  to its
present  experimental upper  limit.  The  effect of  the  CP-odd phase
$\theta$  on $d_e$  is approximately  one to  two orders  of magnitude
smaller than that of $\phi$. The size of $d_e$ increases with
$\tan\beta$ and mass of the lepton $m_\ell$, it is approximatively 
independent of $A_0$ and~$B_0$, but it generically decreases, as 
functions  of the  soft SUSY-breaking parameters $m_0$,  $M_{1/2}$. 

Based  on our  numerical  results, we  have  also derived  approximate
semi-analytical expressions, which differ  from those presented in the
existing  literature  for SUSY  models  realizing  a GUT-scale  seesaw
mechanism.  Specifically, the flavour blind CP-odd phases 
lead to a  scaling of the lepton EDM $d_l  \propto m_l \tan\beta /m_N^{y}$, 
where $2/3<y<1$. Further $d_l$  generally decreases with $M_{SUSY}$, 
but that cannot be described with a simple scaling law. The dependences on SUSY breaking 
parameters $A_0$ and $B_0$ are weak in the largest part of the 
parameter space. The linear dependence on $\tan\beta$ and  the dependence on 
heavy neutrino mass are new results of this paper. In comparison the
$\tan\beta$ depedence in Ref.~\cite{Peskin} is, depending on its 
magnitude, either cubic or constant. Given  the  current 
experimental limits  on $d_e$, we identified a  significant portion of
the $\nu_R$MSSM parameter space with  maximal CP phase $\phi = \pi/2$,
where the electron EDM $d_e$ can have values comparable to the present
and future experimental sensitivities.  The effect of sneutrino-sector
CP  violation  on the  neutron  and Mercury  EDMs  is  expected to  be
suppressed,  which is  a  distinctive  feature for  the  class of  the
$\nu_R$MSSM scenarios studied in this paper.

\subsection*{Acknowledgements}
\vspace{-3mm}
\noindent
The    work    of    AP     is    supported    in    part    by    the
Lancaster--Manchester--Sheffield  Consortium  for Fundamental  Physics
under STFC  grant ST/J000418/1.  AP also  acknowledges partial support
by an IPPP  associateship from Durham University.  The  work of AI and
LP was  supported by  the Ministry of  Science, Sports  and Technology
under contract 119-0982930-1016.

\begin{appendix}

\section{Appendix}\label{f3.4and3.5}

Here we present detailed analytical expressions for all the quantities
that   appear    in   the   formfactors    $G^{L,SB}_{ll\gamma}$   and
$G^{R,SB}_{ll\gamma}$,  given in (\ref{GllgLSB})  and (\ref{GllgRSB}),
respectively.  To start with, the variables $\lambda_X$ are defined as
$\lambda_X=m_X^2/M_W^2$,                  for                 instance
$\lambda_{\tilde{e}}=m^2_{\tilde{e}}/M_W^2$.  The integrals $J^a_{bc}$
derived from loop integrations~\cite{IPP2013} are UV finite. These are
given by
\begin{eqnarray}
  \label{Jabc}
J^a_{bc} &=&
 (-1)^{a-n_b-n_c} 
 \int_0^\infty \frac{dx x^{1+a}}{(x+\lambda_b)^{n_b}(x+\lambda_c)^{n_c}}\ .
\end{eqnarray}
The couplings $\tilde{V}^{0 \ell L}_{l m a}$ and $\tilde{V}^{0 \ell R}_{l m a}$ 
read: 
\begin{eqnarray}
\tilde{V}^{0 \ell L}_{l m a}
 &=&
 -\sqrt{2} t_w Z^*_{m1} (R_R^{\tilde{e}})^*_{al}
 - \frac{(m_e)_l}{\sqrt{2} c_\beta M_W} Z^*_{m3} (R_L^{\tilde{e}})^*_{al}
\\
\tilde{V}^{0 \ell R}_{l m a}
 &=&
\frac{1}{\sqrt{2} c_w}(c_w Z_{m2} +s_w Z_{m1}) (R^{\tilde{e}}_L)^*_{al}
- \frac{(m_e)_l}{\sqrt{2}c_\beta M_W} Z_{m3} (R_R^{\tilde{e}})^*_{al}
\end{eqnarray}
where   $t_w=\tan\theta_w$,   $c_w=\cos\theta_w$,  $s_w=\sin\theta_w$,
$c_\beta=\cos\beta$.  The unitary matrices  ${\cal U}$ and ${\cal V}$,
which diagonalize the chargino mass matrix, and the unitary matrix $Z$
diagonalizing    the     neutralino    mass    matrix     are    taken
from~\cite{DGR_Dress}.    Finally,   the    following   lepton-slepton
disalignment matrices may be defined:
\begin{eqnarray}
R^{\tilde{e}L}_{ak} &=& U^{\tilde{e}}_{ia} U^{e_L*}_{ik}\; ,
\nonumber\\
R^{\tilde{e}R}_{ak} &=& U^{\tilde{e}}_{i+3a} U^{e_R*}_{ik}\; ,
\end{eqnarray}
where  $U^{e_L}$, $U^{e_R}$ and  $U^{\tilde{e}}$ are  unitary matrices
diagonalizing the lepton and slepton mass matrices, with $a=1,\dots,6$
and $i,k=1,2,3$.

\end{appendix}

\end{document}